# Condensation Under Controlled Cooling: A Simulation Study

Barnana Pal
Saha Institute of Nuclear Physics, 1/AF, Bidhannagar, Kolkata-700064.

Abstract:

The formation, growth, structure and cluster size distribution (CSD) properties in a two-dimensional system of particles interacting with Lennard-Jones (LJ) potential under controlled cooling condition have been studied using Monte-Carlo (MC) method considering modified Metropolis algorithm to introduce realistic thermal motion of the particles. The system, initially at relatively higher temperature $T_i$, undergoes temperature reduction following exponential law with decay constant $\alpha$ to a lower temperature $T_f$ and subsequently reaches equilibrium. The equilibrium phase configuration depends strongly on the number density c of particles and $\alpha$. The root mean square particle displacement in the final equilibrium phase shows maximum value for $\alpha = \alpha_c \sim 10^{-3}$ for all c. The CSD properties obtained at $\alpha = 10^{-3}$ shows a sharp peak in the lower cluster size region for low c. The peak shifts towards higher cluster size for lower $\alpha$. The CSD fits well with a modified Gamma distribution function. All of the particles in the system form a single cluster when c is larger than a critical value $c_c$ (~0.5). A compact well-defined ordered structure is obtained for $c \gtrsim c_c$ and $\alpha \ll \alpha_c$.

**I. Introduction.**

In a thermodynamic system at high temperature T, the kinetic energy of any of the constituent particles is large compared to the potential energy of particle-particle interaction and the system is in the gaseous state. As the temperature is reduced, the kinetic energy of individual particles decrease and when its magnitude becomes comparable to the potential energy of particle-particle interaction, particles coming to the close proximity with each other due to their random thermal motion get a chance to be trapped to form clusters. Similar situation arises for growth of crystals from solution or melt. For solution growth processes, the solute molecules come closer as the solvent evaporates and form regular ordered clusters or crystals. The dynamical processes in these cases are strongly dependent on the internal and external parameters. Study of the behaviour of such systems, both from microscopic and macroscopic viewpoint is important due to their application in various fields like formation and growth of crystals [1-3], thin films [4-6], cloud and raindrops [7,8] and many others. Modelling these systems using Monte-Carlo (MC) or Molecular dynamics (MD) simulation [9,10] are also important in understanding these systems. In this connection we propose a simplified model system consisting of p particles confined in a two-dimensional (2D) space and interacting with Lennard-Jones (LJ) potential. The relaxation dynamics of the system under quenching condition has already been reported [11]. The present study is on the dynamical behaviour under controlled cooling condition. Initially the system is assumed to be in a relatively high temperature state in the liquid-like or gas-like phase. The model and the method of simulation are described in Sec.II. Simulation results are presented and discussed in Sec.III and Sec.IV gives the conclusion.

**II. System model and simulation method.**

The model system consists of p point particles confined in a two-dimensional square space with area $\ell \times \ell$ representing a configuration with concentration $c=p/\ell^2$. The particles interact with each other through Lennard-Jones (LJ) potential given by,

$$V(r) = \varepsilon[(r_o/r)^{12} - 2(r_o/r)^6] \quad \ldots\ldots\ldots\ldots\ldots\ldots\ldots\ldots\ldots\ldots (1).$$

$\varepsilon$ is the depth of the potential function and $r_o$ is the position of potential minimum. For simulation, we take $\varepsilon=1$, $r_o=1$ and $\ell=30$. The system temperature T is measured in units of $\varepsilon$, $T \equiv kT/\varepsilon$, k is Boltzmann constant. Initially the system is assumed to be in equilibrium configuration at a temperature $T_i=0.2$. The choice of $T_i$ is based on the fact that the LJ potential function, shown in figure1, has maximum slope at $r/r_o= 1.11$ and at this point $V/\varepsilon \approx -0.8$. Consequently, particles at T~ 0.2 are expected to be in the liquid-like or gas-like phase depending on c and on lowering the temperature, there is possibility for the formation of clusters with ordered structures. To generate the equilibrium configuration at $T_i$, p particles are placed randomly in the 2D space considering hard-core repulsion at short distances in the LJ potential field and 20,000 Monte-Carlo steps (MCS) per particle are performed on this system at $T_i=0.2$ following Metropolis algorithm with the introduction of temperature dependent MCS [11]. T is now allowed to decay exponentially following the relation $T = T_i \exp(-\alpha S)$ with different decay constant $\alpha$ to reach final temperature $T_f = 0.06$, sufficiently low compared to $T_i$. S denotes the number of MCS/particle. On reaching $T_f$, $10^5$ MCS/particle are allowed to ensure equilibrium.

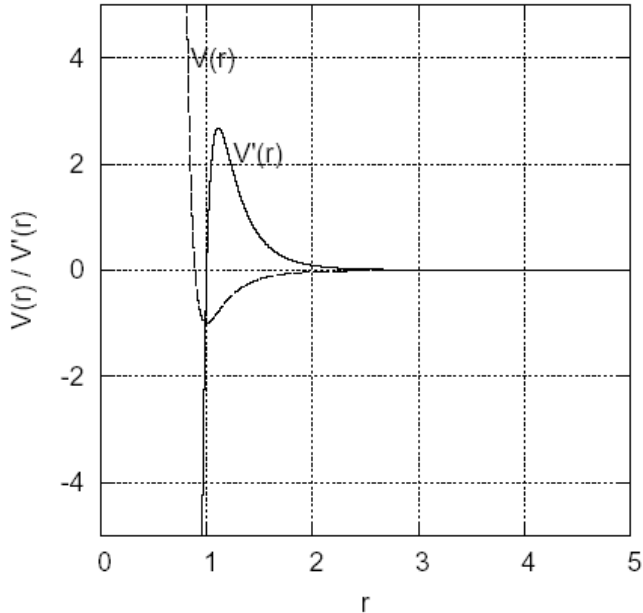

*Figure 1: Normalized Lennard-Jones potential function V(r) (dotted line) and its slope V'(r) (solid line).*

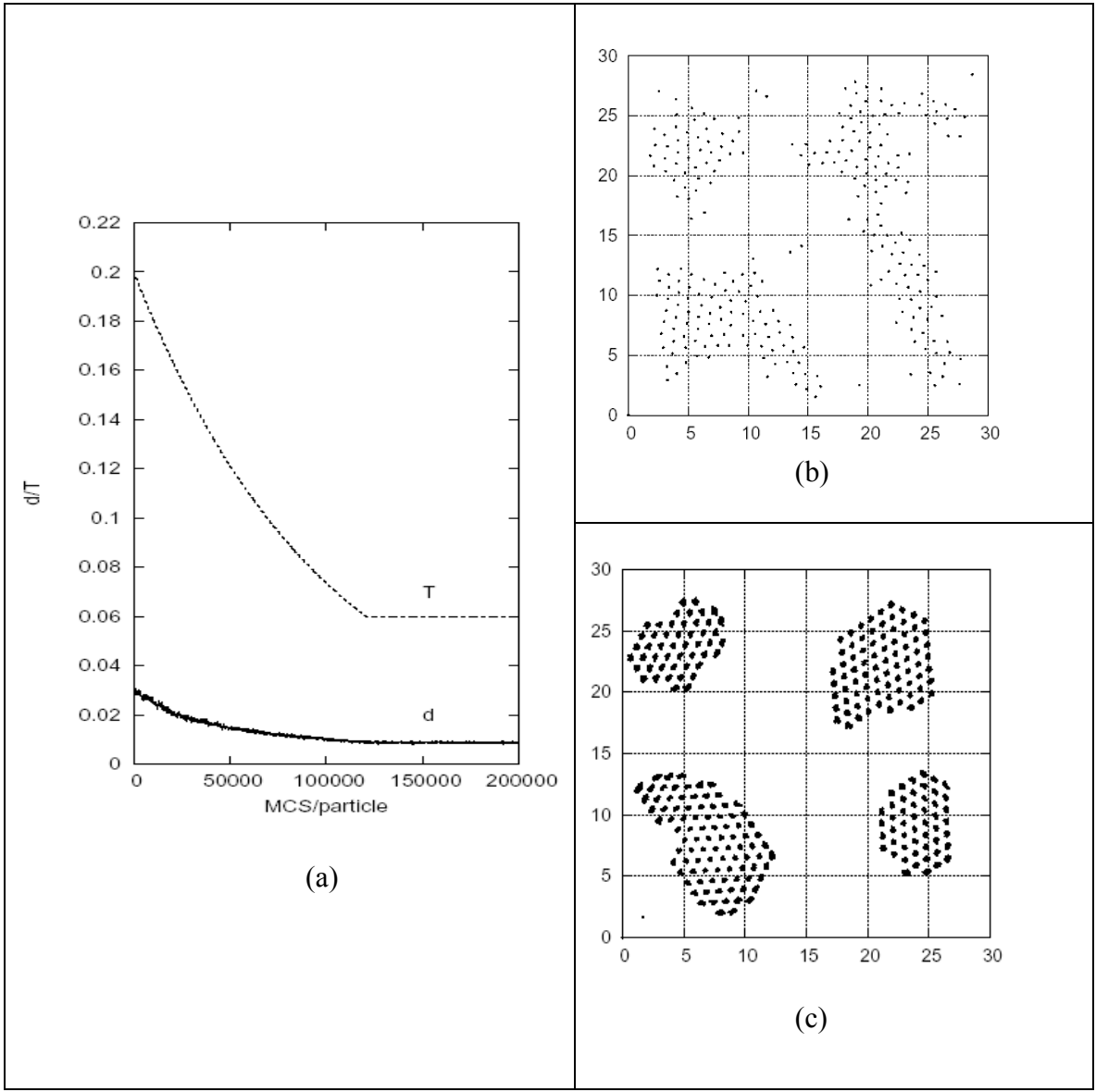

Figure 2: Typical nature of variation of d and T with S for $c=0.4$ and $\alpha = 10^{-5}$ is presented in (a) and (b) & (c) present initial and final equilibrium configuration. Fig1 (c) is the superposition of 50 consecutive snapshots taken at an interval of 100 MCS/particle.

The root mean square deviation d of the particles from their respective previous positions are calculated at an interval of 100 MCS/particle and assuming exponential decay of d with S, final equilibrium value $d_o$ of d is determined by standard curve fitting method. $d_o$ is determined for three to five (for large fluctuation) different initial configurations and average $<d_o>$ is calculated. Variation of $<d_o>$ with $\alpha$ and c has been studied. Fig2(a) presents the typical nature of variation of T and d with S for c=0.4 and $\alpha = 10^{-5}$ and (b) and (c) present initial and final equilibrium configuration. Fig 2(c) is the superposition of 50 consecutive snapshots taken at an interval of 100 MCS/particle and the dot at the left corner shows the size of a single particle. This indicates that in the final equilibrium configuration, the particles in the system are arranged in a regular ordered structure and fluctuates around a mean position due to thermal motion.

To determine the cluster size distribution (CSD) function for various c, similar process is followed to achieve equilibrium and in the equilibrium configuration number of clusters formed with definite sizes are counted. n particles are considered to belong to a particular cluster of normalized size ñ=n/p if each of them have at least one neighbour within a distance $1.5r_o$ [12]. Clusters are counted by dividing the normalized size range 0-1 into several segments of equal span and all the clusters belonging to a particular segment are considered to be of same size equal to the average size of that segment. For low values of c (c≤0.3), 50 configurations are found to be sufficient to determine the CSD function whereas for c>0.3, 300 configurations are generated to determine CSD function.

### III. Results and discussion:

The variation of $<d_o>$ with $\alpha$ for c = 0.2, 0.3, 0.4, 0.5, 0.6, 0.7, and 0.8 is presented in fig 3(a). Figure 3(b) shows the variation of $<d_o>$ with c for $\alpha = 10^{-1}$, $10^{-3}$, $10^{-4}$, $10^{-5}$ and $5\times10^{-6}$. It is seen that $<d_o>$ shows a broad maxima around $\alpha = 10^{-3}$ for all values of c and $<d_o>$ is lower for higher c indicating more compact structure at higher particle density. $\alpha = 10^{-1}$ presents the case of quenching and has been studied extensively earlier [12]. The c-T phase diagram obtained in this study shows that at T=0.2 the system behavior is liquid-like and at T=0.06 it is solid-like. Thus present choice of $T_i$ and $T_f$ indicates that the system is cooled down from liquid-like phase to solid-like phase. We observe that as the temperature is decreased slowly, the particles in proximity with others form clusters and consequently arrange themselves in a regular ordered structure. With slower cooling rate, larger and more ordered compact configuration is reached. This is shown in figure4 for c=0.3. Figure4(a), (b), (c) and (d) present the final configurations reached for $\alpha = 10^{-1}$, $10^{-3}$, $10^{-5}$ and $5\times10^{-6}$ respectively starting from the same initial configuration. Each of these four configurations is obtained from superposition of 50 consecutive snapshots taken at an interval of 100 MCS/particle. Fig. 4(d) obtained for $\alpha=5\times10^{-6}$ shows the formation of two clusters with perfectly ordered structure. Secondly, it is observed that if the number density of particles c is less than a critical value $c_c \approx 0.5$, many clusters are formed randomly depending on the initial configuration but for c ≳ $c_c$ all of the particles in the system join to form a single cluster. Also for fast cooling process, voids or defects are observed in the final equilibrium configuration, whereas for sufficiently slow cooling process, $\alpha \ll \alpha_c$, a compact and perfectly ordered configuration is obtained. This is shown in figure5 for c=0.5. Figure5(a), (b), (c) and (d) are final configurations reached for $\alpha = 10^{-1}, 10^{-3}, 10^{-5}$ and $5 \times 10^{-6}$ starting from the same initial configuration. These are also obtained from superposition of 50 consecutive snapshots taken at an interval of 100 MCS/particle. It is to be noted that a perfectly ordered compact configuration, figure5(d) is obtained for the extremely slow cooling rate $\alpha = 5 \times 10^{-6}$.

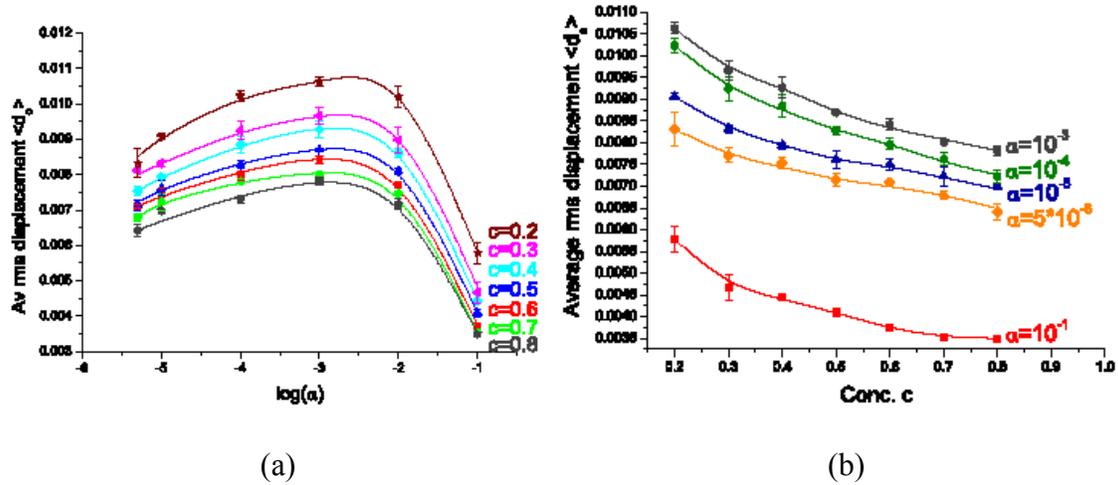

(a)                                      (b)

*Figure 3: The variation of $<d_o>$ with $\alpha$ for $c$ = 0.2, 0.3, 0.4, 0.5, 0.6, 0.7, and 0.8 is presented in (a). (b) shows the variation of $<d_o>$ with c for $\alpha = 10^{-1}, 10^{-3}, 10^{-4}, 10^{-5}$ and $5 \times 10^{-6}$.*

The CSD function for c=0.2, 0.3 and 0.4 are determined for $\alpha = 10^{-3}$ and presented in figure 6(a). The points are obtained from MC simulation and lines give the best-fit result assuming a modified Gamma distribution function [13] of the form

$$f(\tilde{n}) = a\, \tilde{n}^{\mu} \exp(-\lambda \tilde{n}) \quad \ldots \ldots \ldots (2)$$

where $a$, $\mu$ and $\lambda$ are parameters dependent on c. It is clear from fig 6 that the probability for the formation of small size clusters is larger for lower c and faster cooling rate. With the increase of c, probability for the formation of large size cluster increases and for c=0.4 there is a finite probability to form a single cluster. For $c \gtrsim c_c (\approx 0.5)$ probability of formation of a single cluster is 1. Fig 6(b) presents a comparison between CSD functions obtained for $\alpha = 10^{-1}, 10^{-3}$ and $5 \times 10^{-6}$ for c=0.2. It shows that probability of formation of larger clusters increases for slower cooling process.

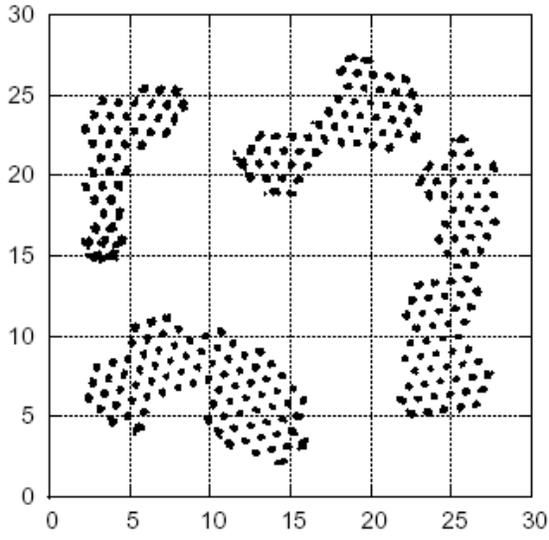 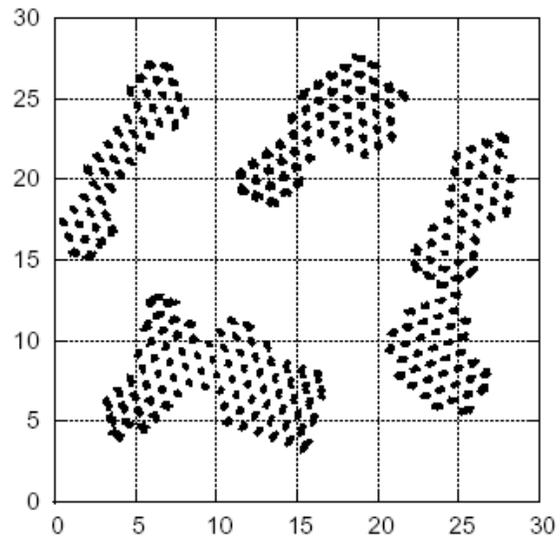

(a)　　　　　　　　　　　　　(b)

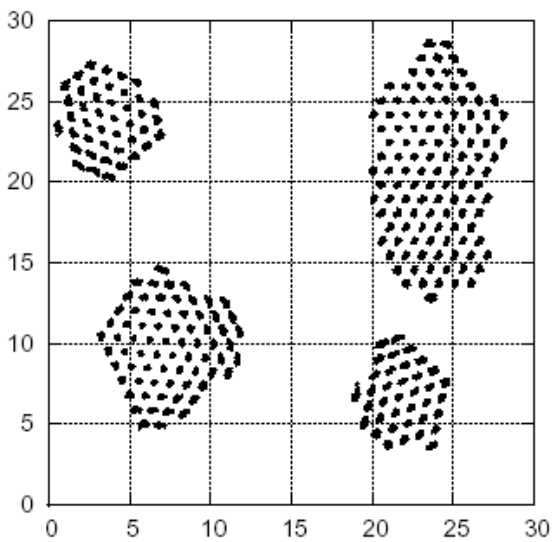 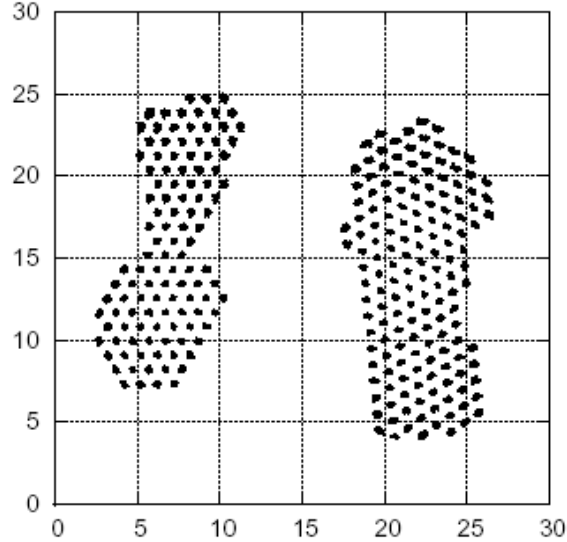

(b)　　　　　　　　　　　　　(d)

*Figure 4: Final equilibrium configurations reached for c = 0.3 starting from the same initial configuration for (a) $\alpha= 10^{-1}$, (b) $\alpha= 10^{-3}$, (c) $\alpha= 10^{-5}$ and (d) $\alpha= 5\times10^{-6}$. Each of these four configurations is obtained from superposition of 50 consecutive snapshots taken at an interval of 100 MCS/particle.*

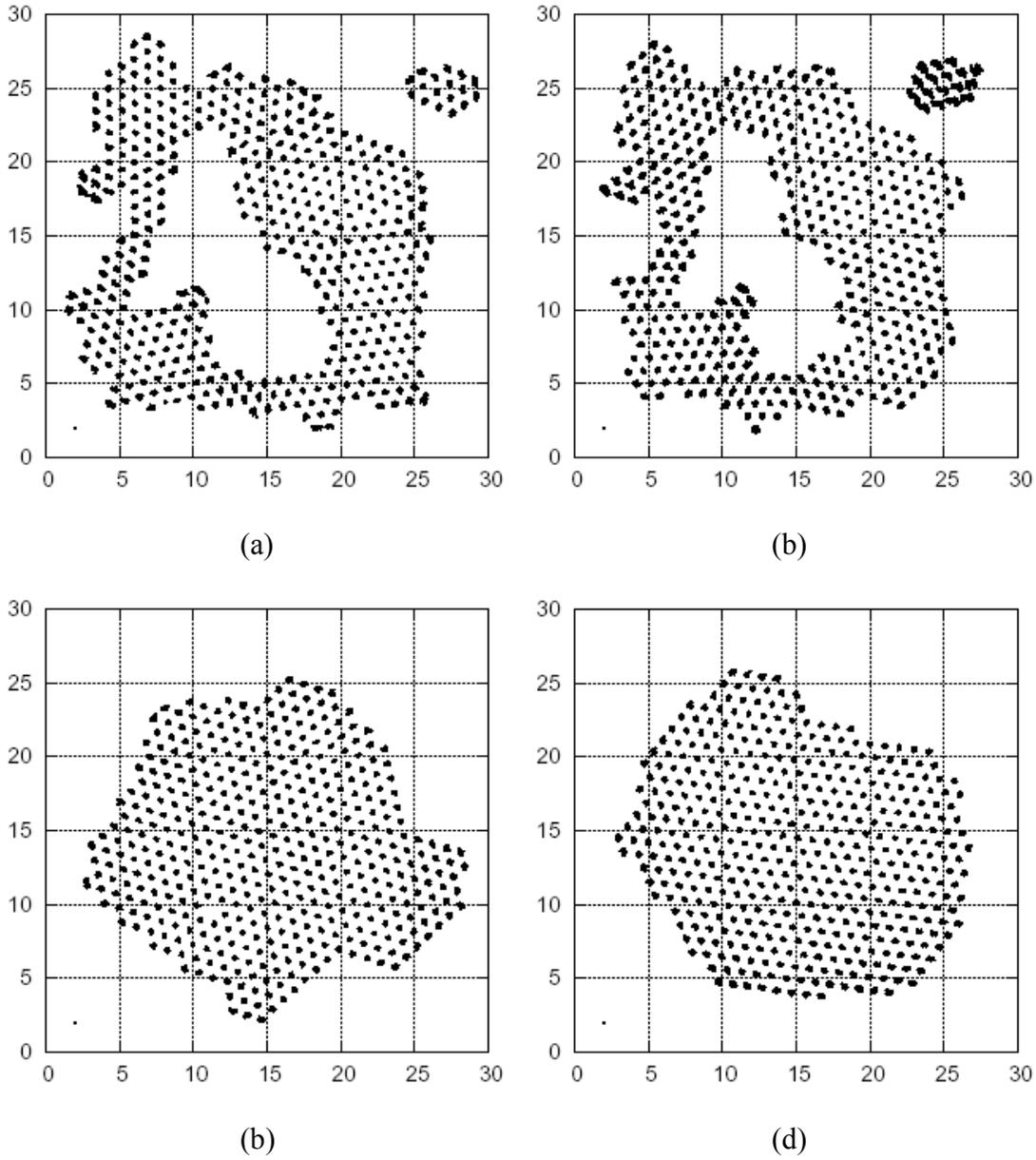

*Figure 5: Final equilibrium configurations reached for c = 0.5 starting from the same initial configuration for (a) $\alpha= 10^{-1}$, (b) $\alpha= 10^{-3}$, (c) $\alpha= 10^{-5}$ and (d) $\alpha= 5\times10^{-6}$. Each of these four configurations is obtained from superposition of 50 consecutive snapshots taken at an interval of 100 MCS/particle and the dot at the left corner shows the size of a single particle.*

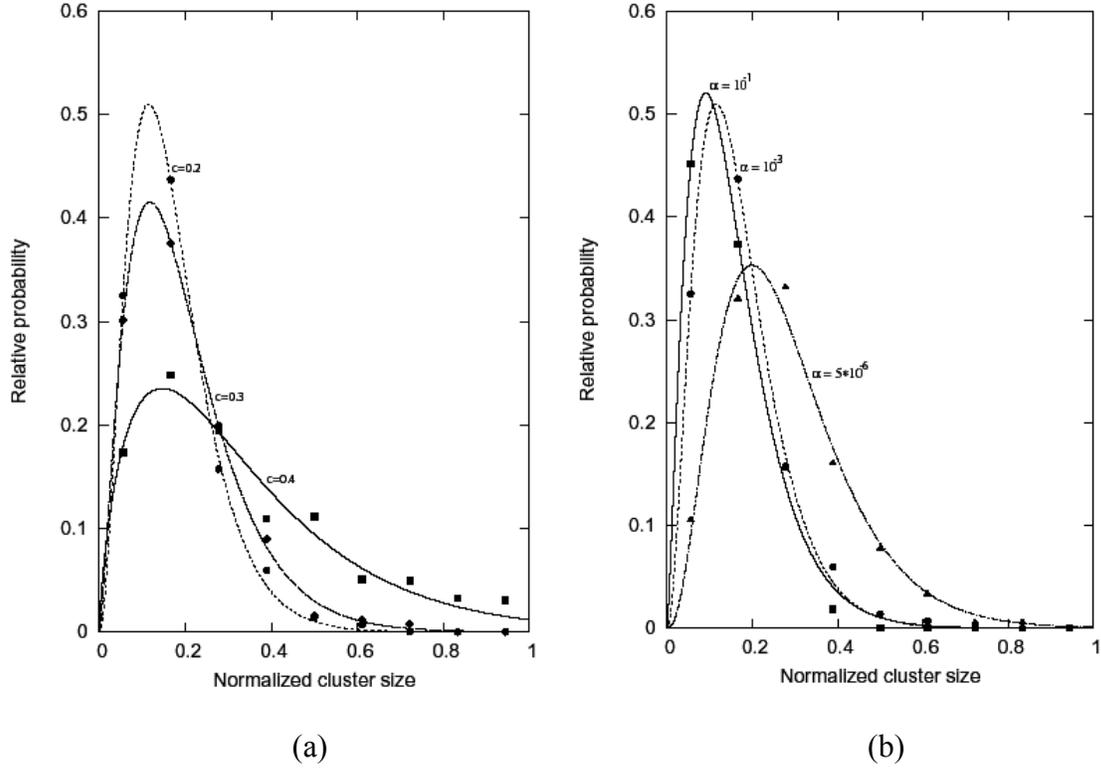

(a)                                                                (b)

*Figure 6: The cluster size distribution (CSD) function for c=0.2, 0.3 and 0.4 obtained at $\alpha = 10^{-3}$ is presented in (a). The points are obtained from MC simulation and lines give the best-fit result assuming a modified Gamma distribution function given by eqn(2). (b) presents a comparison between CSD functions obtained for $\alpha = 10^{-1}$, $10^{-3}$ and $5\times10^{-6}$ for c=0.2.*

## IV. Conclusion:

In the present study, we observe that formation and structure in LJ clusters under slow cooling condition depend on the cooling rate and particle density in the system. For slower cooling process larger size clusters are formed and the CSD function follow a modified form of Gamma distribution with parameters dependent on particle density. There is a critical density $c_c$ ($\approx 0.5$) such that for $c \gtrsim c_c$, all particles in the system join together to form a single cluster. A compact and well-defined ordered structure is obtained for $c \gtrsim c_c$ and $\alpha \ll \alpha_c$.